\documentclass[11pt]{cernrep}
\usepackage{graphicx}
\usepackage{bm}

\begin{document}

\title{Viscous Relativistic Hydrodynamics\footnote{\ \ Work supported 
by the U.S. Department of Energy under contract DE-FG02-01ER41190.}}

\author{Ulrich Heinz}

\institute{Department of Physics, The Ohio State University, Columbus, 
OH 43210, USA}

%

\maketitle

\begin{abstract}
I review recent and not so recent progress on formulating and numerically
implementing a consistent set of relativistic equations which describe the
space-time evolution of viscous relativistic fluids without violating
causality. 
\end{abstract}

\section{Introduction}
%
Ideal fluid dynamics has been used successfully to predict the 
collective flow patterns in Au+Au collisions at the Relativistic 
Heavy-Ion Collider RHIC (for a review see \cite{QGP3}). The ideal
fluid description works well in almost central Au+Au collisions near 
midrapidity at top RHIC energy, but gradually breaks down in more 
peripheral collisions, at forward rapidity, or at lower collision
energies \cite{Heinz:2004ar}. To describe such deviations from ideal
fluid dynamics quantitatively and use the experimental data to extract 
values or phenomenological limits for the transport coefficients of 
the hot and dense matter created during the collision requires the 
numerical implementation of dissipative relativistic fluid dynamics. 
Although a formulation of such a theory which avoids the longstanding 
problems of acausal signal propagation and other instabilities associated
with the original relativistic fluid equations given by Eckart and 
Landau \& Lifshitz, has been known for almost 30 years \cite{IS79},
significant progress towards its numerical implementation has only
been made very recently \cite{Teaney:2004qa,Muronga:2001zk,MR04,CH05}.
At this point, we are only at the very beginning of a program that
will eventually apply a viscous relativistic fluid dynamical approach
to heavy-ion collision data. Existing numerical implementations are
(1+1)-dimensional and can only describe cylindrically symmetric 
transverse expansion with boost-invariant longitudinal dynamics
\cite{MR04,CH05}. The codes are still in the process of being 
tested. In this contribution I try to give a pedagogical summary 
\cite{Rischke} of the theoretical background and summarize a few 
first results.

\section{Ideal fluid dynamics}
%
Before explaining the structure of the equations for causal dissipative
relativistic fluid dynamics, let me quickly review the case of ideal
fluid dynamics. Any fluid dynamical approach starts from the
conservation laws for the conserved charges and for energy-momentum,
\begin{eqnarray}
\label{eq1}
\partial_\mu N^\mu_i &=& 0, \quad i=1,\dots,k,
\\
\label{eq2}
\partial_\mu T^{\mu\nu} &=& 0.
\end{eqnarray}
For simplicity I will restrict myself to $k{\,=\,}1$ (say, $N_\mu=$ 
net baryon number current) and drop the index $i$ in (\ref{eq1}).
It must also ensure the second law of thermodynamics
\begin{equation}
\label{eq3}
\partial_\mu S^\mu \geq 0,
\end{equation}
where $S^\mu$ is the entropy current. Ideal fluid dynamics follows from
these equations under the assumption of local thermal equilibrium, i.e.
if the microscopic collision time scale is very much shorter than any 
macroscopic evolution time scale such that the the underlying phase-space 
distribution $f(x,p)$ relaxes essentially instantaneously to a local
equilibrium form (upper signs for fermions, lower signs for bosons)
\begin{equation}
\label{eq4}
f_{\rm eq}(x,p) = \frac{1}{e^{[p{\cdot}u(x)+\mu(x)]/T(x)}\pm 1},
\end{equation}
where $u^\mu(x)$ is the local fluid velocity at point $x$, $\mu(x)$ is 
the local chemical potential associated with the conserved charge 
$N$ (it enters with opposite sign in the distribution $\bar f$ for
antiparticles), and $T(x)$ is the local temperature. Plugging this into 
the kinetic theory definitions 
\begin{eqnarray}
\label{eq5}
N^\mu(x) &=& \int \frac{d^3p}{E} p^\mu [f(x,p)-\bar f(x,p)],
\\
\label{eq6}
T^{\mu\nu}(x) &=& \int \frac{d^3p}{E} p^\mu p^\nu [f(x,p)+\bar f(x,p)],
\\
\label{eq7}
S^\mu(x) &=& - \int \frac{d^3p}{E} p^\mu 
            \bigl[f(x,p)\ln f(x,p)\pm(1{\mp}f(x,p))\ln(1{\mp}f(x,p)) + 
                 (f\leftrightarrow\bar f)\bigr]
\end{eqnarray}  
leads to the ideal fluid decompositions
\begin{eqnarray}
\label{eq8}
N_{\rm eq}^\mu &=& n\, u^\mu,
\\
\label{eq9}
T_{\rm eq}^{\mu\nu} &=& e\,u^\mu u^\nu - p\, \Delta^{\mu\nu} \quad
(\mbox{with} \ \Delta^{\mu\nu}=g^{\mu\nu}{-}u^\mu u^\nu),
\\
\label{eq10}
S_{\rm eq}^\mu &=& s\, u^\mu,
\end{eqnarray}  
where the local net charge density $n$, energy density $e$, pressure $p$
and entropy density $s$ are given by the standard integrals over
the thermal equilibrium distribution function in the local fluid
rest frame and are related by the fundamental thermodynamic relation
\begin{equation}
\label{eq11}
T\,s = p - \mu\, n + e.
\end{equation}
Inserting Eqs.~(\ref{eq5}-\ref{eq7}) into Eqs.~(\ref{eq1},\ref{eq2}) 
yields the relativistic ideal fluid equations shown in 
Eqs.~(\ref{eq12}-\ref{eq14}) below. Using Eq.~(\ref{eq11}) together with 
the Gibbs-Duhem relation $dp=s\,dT+n\,d\mu$, it is easy to prove that, 
in the absence of shock discontinuities, these equations also conserve 
entropy, i.e. $\partial_\mu S^\mu=0$.

Note that the validity of the decompositions (\ref{eq5}-\ref{eq7}) only
requires local momentum isotropy (i.e. that in the local fluid rest frame 
the phase-space distribution reduces to a function of energy $E$ only, 
$f(x,p)=f(p{\cdot}u(x);T(x),\mu(x))$), but not that the distribution 
function has the specific exponential form (\ref{eq4}) which maximizes 
entropy. This may have relevance in situations where the time scale for 
local momentum isotropization is much shorter than for thermalization 
(i.e. it is much easier to change the direction of the particles'\ momenta
than their energies), with the macroscopic hydrodynamic time scale
in between. In this case the local microscopic states would not 
maximize entropy, and the relation (\ref{eq11}) would not hold between
the quantities $e,\,p,\,n,$ and $s$ defined through 
eqs.~(\ref{eq5}-\ref{eq10}). Still, they would follow ideal fluid
dynamical evolution since entropy production by microscopic kinetic 
energy-shifting processes would happen only on time scales which are
large compared to the macroscopic evolution time scales. 

The ideal fluid equations read (with $\theta\equiv\partial{\cdot}u$ denoting
the local expansion rate)
\begin{eqnarray}
\label{eq12}
\dot n &=& - n\, \theta,
\\
\label{eq13}
\dot e &=& - (e+p)\, \theta,
\\
\label{eq14}
{\dot u}^\mu &=& \frac{\nabla^\mu p}{e+p},
\end{eqnarray}
where we decomposed the partial derivative 
$\partial^\mu=u^\mu D +\nabla^\mu$ into ``longitudinal'' 
and ``transverse'' components $D=u^\nu\partial_\nu$ and 
$\nabla^\mu=\Delta^{\mu\nu}\partial_\nu$ which in the local 
fluid rest frame reduce to the time derivative $\dot f \equiv Df$ and 
spatial gradient. The first two equations describe the dilution
of the local baryon and energy density due to the local
expansion rate $\theta$, while the third describes the 
acceleration of the fluid by the spatial (in the local frame)
pressure gradients, with the enthalpy $e{+}p$ acting as inertia.
The 5 equations (\ref{eq12}-\ref{eq14}) for the 6 unknown functions 
$n,\,e,\,p,\,u^\mu$ (remember that $u^\mu u_\mu=1$) must be closed
by supplying an {\em Equation of State (EOS)} $p=p(n,e)$.

\section{Non-ideal fluid decomposition}
As the hydrodynamic evolution changes the local energy and baryon density,
microscopic process attempt to readjust the local phase-space distribution 
to corresponding new local temperatures and chemical potentials. If this
does not happen sufficiently fast, the phase-space distribution will
start to deviate from its local equilibrium form (\ref{eq4}): 
$f(x,p)=f_{\rm eq}(p{\cdot}u(x);T(x),\mu(x)) + \delta f(x,p)$.
The optimal values for the (readjusted) local temperature and chemical
potential in the first term are fixed by imposing the ``Landau matching 
conditions''
\begin{equation}
\label{eq15}
u_\mu\, \delta T^{\mu\nu} u_\nu = \int \frac{d^3p}{E}\, (u{\cdot}p)^2\,
\delta f(x,p)= u_\mu \,\delta N^\mu =  \int \frac{d^3p}{E} u{\cdot}p \,
\delta f(x,p)=0.
\end{equation}
The remaining deviations $\delta f$ from local equilibrium generate
additional terms in the decompositions of $N^\mu,\,T^{\mu\nu},$ and
$S^\mu$:
\begin{eqnarray}
\label{eq16}
N^\mu &=& N^\mu_{\rm eq} +\delta N^\mu = n\,u^\mu + V^\mu,
\\
\label{eq17}
T^{\mu\nu} &=& T_{\rm eq}^{\mu\nu} + \delta T^{\mu\nu}
           = e\,u^\mu u^\nu -(p+\Pi)\Delta^{\mu\nu}
               + \pi^{\mu\nu} + W^\mu u^\nu +W^\nu u^\mu,
\\
\label{eq18}
S^\mu &=& S^\mu_{\rm eq} +\delta S^\mu = n\,u^\mu + \Phi^\mu.
\end{eqnarray}
The new terms describe a baryon number flow $V^\mu=\Delta^{\mu\nu} N_\nu$ 
in the local rest frame, an energy flow $W^\mu = \frac{e{+}p}{n} V^\mu
+ q^\mu$ (where $q^\mu$ is the ``heat flow vector'') in the local
rest frame, the viscous bulk pressure 
$\Pi = -\frac{1}{3}\Delta_{\mu\nu}T^{\mu\nu}-p$ (which contributes
to the trace of the energy momentum tensor), the traceless viscous
shear pressure tensor $\pi^{\mu\nu}= T^{\langle\mu\nu\rangle} 
\equiv \left[\frac{1}{2}\left(\Delta^{\mu\sigma}
\Delta^{\nu\tau}{+}\Delta^{\nu\sigma}\Delta^{\mu\tau}\right)-\frac{1}{3}
\Delta^{\mu\nu}\Delta^{\sigma\tau}\right]T_{\tau\sigma}$
(where the expression $\langle\mu\nu\rangle$ is a shorthand
for ``traceless and transverse to $u_\mu$ and $u_\nu$'', as defined
by the projector in square brackets), and an entropy flow vector
$\Phi^\mu$ in the local rest frame.

Inserting the decompositions (\ref{eq15},\ref{eq17}) into the conservation
laws (\ref{eq1},\ref{eq2}) yields the {\em non-ideal fluid equations}
\begin{eqnarray}
\label{eq19}
\dot n &=& - n\, \theta - \nabla{\cdot}V +V{\cdot}{\dot u},
\\
\label{eq20}
\dot e &=& - (e{+}p{+}\Pi)\, \theta 
           +\pi_{\mu\nu}\nabla^{\left\langle\mu\right.}
                             u^{\left.\nu\right\rangle}
           -\nabla{\cdot}W + 2\, W{\cdot}{\dot u},
\\
\label{eq21}
(e{+}p{+}\Pi)\,{\dot u}^\mu &=& \nabla^\mu(p{+}\Pi) 
                             -\Delta^{\mu\nu}\nabla^\sigma\pi_{\nu\sigma}
             +\pi^{\mu\nu}{\dot u}_\nu
             -\left[\Delta^{\mu\nu}{\dot W}_\nu + W^\mu\theta 
             +(W{\cdot}\nabla)u^\mu\right].
\end{eqnarray}

The matching conditions (\ref{eq15}) leave the choice of the local 
rest frame velocity $u^\mu$ ambiguous. This ambiguity can be used to
eliminate either $V^\mu$ from Eq.~(\ref{eq16}) (``Eckart frame'', no baryon
flow in the local rest frame), in which case the energy flow reduces to 
the heat flow vector $W^\mu{\,=\,}q^\mu$, or $W^\mu$ from Eq.~(\ref{eq17})
(``Landau frame'', no energy flow in the local rest frame), in which 
case there is a non-zero baryon flow $V^\mu{\,=\,}-\frac{n}{e{+}p}q^\mu$
due to heat conduction in the local rest frame. (Intermediate frames are
also possible, but yield no practical advantage.) For systems with
vanishing net baryon number (as approximately realized in RHIC collisions)
the Eckart frame is ill-defined, so we will use the Landau frame.
In this frame, for baryon-free systems with $n{\,=\,}0$ and no heat 
conduction, the non-ideal fluid equations (\ref{eq19}-\ref{eq21}) simplify 
to
\begin{eqnarray}
\label{eq22}
\dot e &=& - (e{+}p{+}\Pi)\, \theta 
           +\pi_{\mu\nu}\nabla^{\left\langle\mu\right.}
                             u^{\left.\nu\right\rangle},
\\
\label{eq23}
(e{+}p{+}\Pi)\,{\dot u}^\mu &=& \nabla^\mu(p{+}\Pi) 
                             -\Delta^{\mu\nu}\nabla^\sigma\pi_{\nu\sigma}
             +\pi^{\mu\nu}{\dot u}_\nu.
\end{eqnarray}

The non-equilibrium decompositions (\ref{eq16}-\ref{eq18}) involve
1+3+5=9 additional dynamical quantities, the ``dissipative flows''
$\Pi,\,q^\mu$, and $\pi^{\mu\nu}$ (the counting reflects their 
transversality to $u^\mu$ and the tracelessness of $\pi^{\mu\nu}$). 
This means that we need 9 additional dynamical equations which 
should be compatible with the underlying transport theory for the 
non-equilibrium deviation $\delta f(x,p)$. For the baryon-free case
without heat conduction, the number of needed additional equations
reduces to 6.

\section{Kinetic equations for the dissipative flows}
The key property of the kinetic equation governing the evolution of
the phase-space distribution function $f{\,=\,}f_{\rm eq}{+}\delta f$
is that the collision term satisfies the second law of thermodynamics
(\ref{eq3}), i.e. entropy is produced until the system has reached a 
new state of local thermal equilibrium. We don't want to solve the
kinetic theory; instead, we want to write down a phenomenological 
macroscopic theory which is consistent with the constraints arising from
the underlying kinetic theory, in particular the 2$^{\rm nd}$ law. The
macroscopic theory will be constructed from an expansion of the
entropy production rate in terms of the dissipative flows which 
themselves are proportional to the off-equilibrium deviation $\delta f$
of the phase-space distribution. Assuming the latter to be small,
$|\delta f|{\,\ll\,}|f_{\rm eq}|$, this expansion will be truncated
at some low order in the dissipative flows $\delta N^\mu,\,\delta
T^{\mu\nu}$. The expansion will involve phenomenological expansion 
coefficients which, in principle, should be matched to the kinetic 
theory \cite{IS79}. In practice, they will often be considered as
phenomenological parameters to be adjusted to experimental data.
In the end, the extracted values must then be checked for consistency 
with the entire approach, by making sure that the dissipative corrections
are indeed sufficiently small to justify truncation of the expansion 
{\em a posteriori}.

The equilibrium identity (\ref{eq11}) can be rewritten as
\begin{equation}
\label{eq24}
S_{\rm eq}^\mu = p(\alpha,\beta)\beta^\mu - \alpha N_{\rm eq}^\mu
+\beta_\nu T_{\rm eq}^{\nu\mu},
\end{equation}
where $\alpha{\,\equiv\,}\frac{\mu}{T},\,\beta{\,\equiv\,}\frac{1}{T},$ and 
$\beta_\nu{\,\equiv\,}\frac{u_\nu}{T}$. The most general off-equilibrium
generalization of this is \cite{IS79}
\begin{equation}
\label{eq25}
S^\mu \equiv S_{\rm eq}^\mu + \Phi^\mu = p(\alpha,\beta)\beta^\mu - 
\alpha N^\mu +\beta_\nu T^{\nu\mu} + Q^\mu(\delta N^\mu,\delta T^{\mu\nu}),
\end{equation}
where, in addition to the first order contributions implicit in
the second and third terms of the r.h.s., $Q^\mu$ includes terms which
are second and higher order in the dissipative flows $\delta N^\mu$
and $\delta T^{\mu\nu}$. [Note that, by using the identity (\ref{eq11})
between the equilibrium quantities, Eq.~(\ref{eq25}) can be written
in the simpler-looking form $S^\mu{\,=\,}s\,u^\mu{+}\frac{q^\mu}{T}{+}Q^\mu$
but this is not helpful for calculating the entropy production rate.]

The form of the expansion (\ref{eq25}) is constrained by the
2$^{\rm nd}$ law $\partial_\mu S^\mu{\,\geq\,}0$. To evaluate this 
constraint it is useful to rewrite the Gibbs-Duhem relation 
$dp{\,=\,}s\,dT+n\,d\mu$ as
\begin{equation}
\label{eq26}
\partial_\mu\left( p(\alpha,\beta)\beta^\mu\right) = 
N_{\rm eq}^\mu\partial_\mu\alpha - T_{\rm eq}^{\mu\nu}\partial_\mu\beta_\nu.
\end{equation}
With additional help from the conservation laws (\ref{eq1},\ref{eq2}), the 
entropy production then becomes
\begin{equation}
\label{eq27}
\partial_\mu S^\mu = - \delta N^\mu\partial_\mu\alpha
+ \delta T^{\mu\nu}\partial_\mu\beta_\nu +\partial_\mu Q^\mu.
\end{equation}
Using Eqs.~(\ref{eq16},\ref{eq17}) to express $\delta N^\mu$ and $\delta 
T^{\mu\nu}$ in terms of the scalar, vector and tensor dissipative flows
$\Pi,\,q^\mu,$ and $\pi^{\mu\nu}$, and introducing corresponding scalar,
vector and tensor thermodynamic forces (in terms of gradients of the
thermodynamic equilibrium variables) which drive these dissipative flows,
$X{\,\equiv\,}{-}\theta{\,=\,}{-}\nabla{\cdot}u$, \ 
$X^\nu{\,\equiv\,}\frac{\nabla^\nu T}{T}-{\dot u}^\nu{\,=\,}-\frac{nT}{e{+}p}\,
\nabla^\nu\!\left(\frac{\mu}{T}\right)$, and
$X^{\mu\nu}{\,\equiv\,}\nabla^{\left\langle\mu\right.}
 u^{\left.\nu\right\rangle}$ (note that $X^{\mu\nu}{=}X^{\langle\mu\nu\rangle}$
is traceless and transverse to $u$), the 2$^{\rm nd}$ law constraint 
can be further recast into
\begin{equation}
\label{eq28}
 T \partial_\mu S^\mu = \Pi X - q^\mu X_\mu +\pi^{\mu\nu} X_{\mu\nu}
+ T \partial_\mu Q^\mu \geq 0.
\end{equation}
Note that the first three terms on the r.h.s. are first order while the 
last term is higher order in the dissipative flows.

\subsection{Standard dissipative fluid dynamics (first order theory)}
%
The standard approach (which can be found, for example, in the 
{\em Fluid Dynamics} volume of Landau \& Lifshitz) one neglects the higher
order contributions and sets $Q^\mu{\,=\,}0$. The inequality (\ref{eq28})
can than be always satisfied by postulating linear relationships
between the dissipative flows and the thermodynamic forces,
\begin{eqnarray}
\label{eq29}
  \Pi=-\zeta\theta,\qquad
  q^\nu=-\lambda \frac{nT^2}{e{+}p}\,\nabla^\nu\!\left(\frac{\mu}{T}\right),
  \qquad \pi^{\mu\nu} = 2\, \eta\, \nabla^{\left\langle\mu\right.}
  u^{\left.\nu\right\rangle},
\end{eqnarray}
with positive {\em transport coefficients} $\zeta{\,\geq\,}0$ ({\em bulk
viscosity}),  $\lambda{\,\geq\,}0$ ({\em heat conductivity}), and 
$\eta{\,\geq\,}0$ ({\em shear viscosity}):
\begin{equation}
\label{eq30}
T \partial{\cdot}S = \frac{\Pi^2}{\zeta} - \frac{q^\alpha q_\alpha}{2\lambda T}
+ \frac{\pi^{\alpha\beta}\pi_{\alpha\beta}}{2\eta}\geq 0.
\end{equation}
(The minus sign in front of the second term is necessary because $q^\mu$,
being orthogonal to $u^\mu$, is spacelike, $q^2<0$.) These are the desired 
9 equations for the dissipative flows. 

Unfortunately, using these relations in the hydrodynamic equations 
(\ref{eq19}-\ref{eq21}) leads to hydrodynamic evolution with acausal 
signal propagation: if in a given fluid cell at a certain time a 
thermodynamic force happens to vanish, the corresponding dissipative
flow also stops instantaneously. This contradicts the fact that the 
flows result from the forces through microscopic scattering which 
involves relaxation on a finite albeit short kinetic time scale.
To avoid this type of acausal behaviour one must keep $Q^\mu$.

\subsection{Second order Israel-Stewart theory}
%
A causal theory of dissipative relativistic fluid dynamics is obtained
by keeping $Q^\mu$ up to terms which are second order in the irreversible 
flows. For simplicity I here consider only the baryon-free case 
$n{\,=\,}q^\mu{\,=\,}0$, for a general treatment see 
\cite{IS79,Muronga:2001zk}. One writes \cite{IS79}
\begin{equation}
\label{eq31}
  Q^\mu = -\left(\beta_0\Pi^2 + \beta_2\pi_{\nu\lambda}\pi^{\nu\lambda}\right)
  \frac{u^\mu}{2T}
\end{equation}
(with phenomenological expansion coefficients $\beta_0,\,\beta_2$) and 
computes (after some algebra using similar techniques as before) the
entropy production rate as
\begin{equation}
\label{eq32}
  T \partial{\cdot}S = \Pi \left[ -\theta -\beta_0 \dot\Pi - \Pi T 
  \partial_\mu\left(\frac{\beta_0 u^\mu}{2T}\right)\right]
  + \pi^{\alpha\beta}\left[\nabla_{\left\langle\alpha\right.}
  u_{\left.\beta\right\rangle} - \beta_2{\dot\pi}_{\alpha\beta}
  -\pi_{\alpha\beta} T \partial_\mu\left(\frac{\beta_2 u^\mu}{2T}\right)
  \right].
\end{equation}
From the expressions in the square brackets we see that the thermodynamic 
forces $-\theta$ and $\nabla_{\left\langle\alpha\right.} 
u_{\left.\beta\right\rangle}$ are now self-consistently modified by
terms involving the time derivatives (in the local rest frame) of the 
irreversible flows $\Pi$, $\pi_{\alpha\beta}$. This leads to dynamical
(``transport'') equations for the latter. We can ensure the 2$^{\rm nd}$ 
law of thermodynamics by again writing the entropy production rate in 
the form (\ref{eq30}) (without the middle term), which amounts to 
postulating 
\begin{eqnarray}
\label{eq33}
\dot \Pi &=& -\frac{1}{\tau_{_\Pi}}\left[ \Pi +\zeta \theta +
\Pi\zeta T \partial_\mu\left(\frac{\tau_{_\Pi} u^\mu}{2\zeta T}\right)\right]
\approx -\frac{1}{\tau_{_\Pi}}\bigl[ \Pi +\zeta \theta\bigr],
\\
\label{eq34}
\dot \pi_{\alpha\beta} &=& -\frac{1}{\tau_\pi}\left[ \pi_{\alpha\beta}
- 2 \eta \nabla_{\left\langle\alpha\right.} u_{\left.\beta\right\rangle}
+ \pi_{\alpha\beta} \eta T \partial_\mu\left(\frac{\tau_\pi u^\mu}{2\eta T}
\right)\right]
\approx -\frac{1}{\tau_\pi}\left[ \pi_{\alpha\beta}
   - 2 \eta \nabla_{\left\langle\alpha\right.} u_{\left.\beta\right\rangle}
\right].
\end{eqnarray}
Here I replaced the coefficients $\beta_{0,2}$ by the relaxation times 
$\tau_{_\Pi}{\,\equiv\,}\zeta\beta_0$ and $\tau_\pi{\,\equiv\,}2\eta\beta_2$.
In principle both $\zeta,\eta$ and $\tau_{_\Pi},\tau_\pi$ should be calculated
from the underlying kinetic theory. We will use them as phenomenological
parameters, noting that for consistency the microscopic relaxation rates
should be much larger than the local hydrodynamic expansion rate, 
$\tau_{_{\pi,\Pi}}\theta{\,\ll\,}1$.

Let me shortly comment about the approximation in the second equalities
in Eqs.~(\ref{eq33},\ref{eq34}): We are using an expansion scheme for the
entropy production rate in which the thermodynamic forces and irreversible
flows are assumed to be small perturbations. The approximation in
Eqs.~(\ref{eq33},\ref{eq34}) neglects terms which are products of
the irreversible flows with gradients of the thermodynamic equilibrium
quantities which are of the same order as the thermodynamic forces.
These terms are thus effectively of second order in small quantities
and should, for consistency, be neglected relative to the other terms
in the square brackets which are of first order. If one wants to keep
them (as done by Muronga \cite{Muronga:2001zk,MR04}), one should also
keep third-order terms in the entropy flow vector $Q^\mu$ for consistency.
Of course, where the thermodynamic forces and irreversible flows are 
really small, it shouldn't matter whether we keep or drop these terms.
In practice, however, one will use this approach when dissipative effects
are expected to be significant, and the dropped terms may not be extremely
small. In this case I believe that dropping them is more consistent
than keeping them.

There is another reason for dropping these terms: without them, 
Eqs.~(\ref{eq33},\ref{eq34}) are relaxation equations which describe
(in the local rest frame) exponential relaxation (on the time scales 
$\tau_{_{\pi,\Pi}}$) of the irreversible flows to the values given
by Eqs.~(\ref{eq29}) in the first order theory. However, if these 
terms are kept, one has instead equations of the form
\begin{equation}
\label{eq35}
\dot \Pi = -\frac{1}{\tau_{_\Pi}}\Bigl[ \Pi +\zeta \theta +
\Pi\zeta \gamma_{_\Pi}\Bigr] = -\frac{1{+}\gamma_{_\Pi}\zeta}{\tau_{_\Pi}}
\left[ \Pi + \frac{\zeta}{1{+}\gamma_{_\Pi}\zeta}\,\theta\right]
= -\frac{1}{\tau'_{_\Pi}}\Bigl[ \Pi +\zeta'\, \theta\Bigr],
\end{equation}
and similarly for the shear pressure tensor. One sees that both the 
kinetic relaxation time and the viscosity are modified by the
factor $\gamma_{_\Pi}{\,=\,}T \partial_\mu\left(\frac{\tau_{_\Pi} 
u^\mu}{2\zeta T}\right)$ which involves the macroscopic expansion 
rate $\partial_\mu u^\mu$.
This violates the intuition that these transport coefficients should
be expressible through integrals of the kinetic collision term
which involve only microscopic physics (cross sections, local densities, 
etc.) 

In the second order Israel-Stewart formalism, one thus solves the
dissipative hydrodynamic equations (\ref{eq19}-\ref{eq21}) simultaneously
with the kinetic relaxation equations (\ref{eq33},\ref{eq34}) for
the irreversible flows. Let us now look at these equations in more 
detail when expressed in a global coordinate system (and not in local
rest frame coordinates as done up to now).

\section{Transverse expansion dynamics in central collisions at high
energy}
I will restrict my discussion here to azimuthally symmetric systems 
with longitudinal boost invariance. With this approximation we can 
describe the transverse expansion in central collisions at very high 
energy in a domain near midrapidity. A generalized discussion which
relaxes the assumption of azimuthal symmetry and thus allows for
collisions at any impact parameter has been recently given in 
\cite{2dvisc}. 
 
Boost-invariant and azimuthally symmetric systems are conveniently 
described in $(\tau,r,\phi,\eta)$ coordinates where 
$\tau{\,=\,}\sqrt{t^2{-}z^2}$ is longitudinal proper time,
$\eta{\,=\,}\frac{1}{2}\ln[(t{+}z)/(t{-}z)]$ is space-time rapidity, and 
$\bm{r}{\,=\,}(r,\phi)$ are polar coordinates in the plane transverse 
to the beam direction $z$. Such systems are then characterized by 
macroscopic observables which are independent of $\phi$ and $\eta$,
and by azimuthally constant phase-space distributions which depend 
only on the difference $Y{-}\eta$ 
(where $Y{\,=\,}\frac{1}{2}\ln[(E{+}p_z)/(E{-}p_z)]$ is the
momentum-space rapidity of a particle with longitudinal momentum $p_z$
and energy $E$). We use lowercase latin letters to denote vector and 
tensor components in this curvilinear space-time coordinate system. 
The metric tensor in this coordinate system reads
\begin{equation}
\label{eq42}
g^{mn}={\rm diag}\,\bigl(1,-1,-1/r^2,-1/\tau^2\bigr),
\qquad
g_{mn}={\rm diag}\,\bigl(1,-1,-r^2,-\tau^2\bigr).
\end{equation}
The flow velocity is parametrized as 
\begin{equation}
\label{eq44}
u^m = \gamma_r \bigl(1,v_r,0,0\bigr) \quad \mbox{with} \quad
      \gamma_r = \frac{1}{\sqrt{1-v_r^2}} 
\end{equation}
with radial transverse flow velocity $\bm{v}_\perp{\,=\,}v_r(\tau,r)\,\bm{e}_r$
and vanishing flow components $u^\phi$ and $u^\eta$. For vectors and 
tensors, the usual Cartesian derivatives $\partial_\mu$ must be replaced 
by covariant derivatives, denoted by semicolons:
\begin{equation}
\label{eq38}
\partial_\mu j^\nu \to j^n_{\ ;m} = \partial_m j^n
+ \Gamma^n_{mk} j^k, \qquad
\partial_\mu T^{\nu\lambda} \to T^{nl}_{\ \ ;m} =
\partial_m T^{nl} + \Gamma^n_{mk} T^{kl} + T^{nk} \Gamma^l_{km},
\end{equation}
where $\Gamma^i_{jk}{\,=\,}\frac{1}{2}g^{im}\bigl(\partial_j 
g_{km}{+}\partial_k g_{mj} -\partial_m g_{jk}\bigr)$ are
the Christoffel symbols. The nonvanishing components
of $\Gamma^i_{jk}$ are
\begin{equation}
\label{eq39}
\Gamma^\eta_{\eta\tau} = \Gamma^\eta_{\tau\eta} = \frac{1}{\tau},
\quad  \Gamma^\tau_{\eta\eta} = \tau,
\quad  \Gamma^\phi_{\phi r} = \Gamma^\phi_{r\phi} = +\frac{1}{r},
\quad  \Gamma^r_{\phi\phi} = - r.
\end{equation}
The time derivative in the local comoving frame and the local
expansion rate are thus computed as
\begin{eqnarray}
\label{eq45}
&&D = u\cdot\partial = \gamma_r \bigl(\partial_\tau + v_r \partial_r\bigr), 
\\
\label{eq46}
&&\theta = \partial\cdot u 
        = \frac{1}{\tau}\, \partial_\tau\left(\tau\gamma_r\right) 
        + \frac{1}{r}\, \partial_r\left(r v_r \gamma_r\right) 
\end{eqnarray}

Due to azimuthal symmetry and longitudinal boost invariance, the
$n{\,=\,}\phi$ and $n{\,=\,}\eta$ components of the equations of motion
${T^{mn}}_{;m}{\,=\,}0$ are redundant. The $n{\,=\,}\tau$ and 
$n{\,=\,}r$ components can be written as \cite{MR04,2dvisc}
\begin{eqnarray}
\label{eq47}
&&\frac{1}{\tau}\partial_\tau\Bigl(\tau T^{\tau\tau}\Bigr) +
  \frac{1}{r}\partial_r\Bigl(r T^{\tau r}\Bigr) =
  -\,\frac{p+\Pi+\tau^2 \pi^{\eta\eta}}{\tau},
\\
\label{eq48}
&&\frac{1}{\tau}\partial_\tau\Bigl(\tau T^{\tau r}\Bigr) +
  \frac{1}{r}\partial_r\Bigl(r (T^{\tau r} v_r + {\cal P}_r)\Bigr) =
  +\,\frac{p+\Pi+r^2\pi^{\phi\phi}}{r}.
\end{eqnarray}
With the shorthand notations $\tilde T^{mn}{\,=\,}r \tau T^{mn}$,
$\tilde{\cal P}_r{\,=\,}r \tau {\cal P}_r$, and 
$\tilde v_r{\,=\,}\frac{\tilde T^{\tau r}}{\tilde T^{\tau\tau}}{\,=\,}\frac
{T^{\tau r}}{T^{\tau\tau}}$ these are
brought into ``standard (Cartesian) form''
\begin{eqnarray}
\label{eq49}
&&\partial_\tau \tilde T^{\tau\tau} 
+ \partial_r(\tilde v_r \tilde T^{\tau\tau}) 
= -r \Bigl(p+\Pi+\tau^2 \pi^{\eta\eta}\Bigr),
\\
\label{eq50}
&&\partial_\tau \tilde T^{\tau r} + \partial_r\Bigl(v_r \tilde T^{\tau r} 
  + \tilde {\cal P}_r\Bigr) = +\tau\Bigl(p+\Pi+r^2\pi^{\phi\phi}\Bigr).
\end{eqnarray}
The corresponding transport equations for the dissipative fluxes read
\begin{eqnarray}
\label{eq51}
&&\Bigl(\partial_\tau + v_r\partial_r\Bigr)\pi^{\eta\eta} = 
  -\frac{1}{\gamma_r\tau_\pi}\left[\pi^{\eta\eta}
  -\frac{2\eta}{\tau^2}\left(\frac{\theta}{3}-\frac{\gamma_r}{\tau}
   \right)\right],
\\
\label{eq52}
&&\Bigl(\partial_\tau + v_r\partial_r\Bigr)\pi^{\phi\phi} = 
  -\frac{1}{\gamma_r\tau_\pi}\left[\pi^{\phi\phi}
  -\frac{2\eta}{r^2}\left(\frac{\theta}{3}-\frac{\gamma_r v_r}{r}
   \right)\right],
\\
\label{eq53}
&&\Bigl(\partial_\tau + v_r\partial_r\Bigr)\Pi = 
  -\frac{1}{\gamma_r\tau_{_\Pi}}\left[\Pi + \zeta\theta\right],
\end{eqnarray}
with the following explicit expressions for the shear tensor components:
\begin{eqnarray}
\label{A10}
&&\sigma^{\eta\eta} = \frac{1}{\tau^2} \left(\frac{\theta}{3}
                     -\frac{\gamma_r}{\tau}\right),
\\
\label{A11}
&&\sigma^{\phi\phi} = \frac{1}{r^2} \left(\frac{\theta}{3}
                     -\frac{\gamma_r v_r}{r}\right).
\end{eqnarray}

The hydrodynamic equations require the equation of state (EOS) $p(e)$
for closure, i.e. after each transport step in time we must extract
at each spatial grid point the boost velocity $v_r$ between the global
and local rest frames and the local energy density $e$ from the dynamical
variables $T^{\tau\tau}$ and $T^{\tau r}$. The energy density is obtained 
from
\begin{equation}
\label{eq54}
e = T^{\tau\tau} - v_r T^{\tau r},
\end{equation}
where the radial velocity $v_r$ must be extracted from the implicit
equation 
\begin{equation}
\label{eq55}
v_r = \frac{T^{\tau r}}{T^{\tau\tau}+p(e{=}T^{\tau\tau}{-}v_r T^{\tau r})
      + \Pi - r^2 \pi^{\phi\phi} - \tau^2 \pi^{\eta\eta}}
\end{equation}
by a one-dimensional zero search. This is still the same degree of 
numerical complexity as in the ideal fluid case \cite{Rischke};
for dissipative hydrodynamics {\it without} azimuthal symmetry, however, 
this part of the problem becomes numerically more involved \cite{2dvisc}.

\section{First numerical results}

I close this talk by showing some preliminary results \cite{CH05}
from a numerical simulation of the equations derived in the preceding 
section, using a simple massless ideal gas EOS, 
$p{\,=\,}\frac{1}{3}e$, with $e{\,=\,}aT^4$, 
$a{\,=\,}(16+\frac{21}{2} N_f)\frac{\pi^2}{30}$. We neglect bulk 
viscosity, $\zeta{\,=\,}0$.
 
In classical kinetic theory, explicit expressions can be obtained for the 
viscosity coefficient $\eta$ and the relaxation time $\tau_\pi$ in terms 
of the collision term. For a strongly coupled QGP, neither $\eta$ or 
$\tau_\pi$ are known. We treat them as phenomenological parameters. For 
guidance, we use perturbative \cite{ba90,ar00} and AdS/CFT-based 
\cite{po01} estimates for $\eta$, respectively, and a kinetic theory 
estimate \cite{IS79} for $\tau_\pi$.

The shear viscosity coefficient $\eta$ for hot QCD was determined 
perturbatively to leading logarithmic accuracy in \cite{ba90,ar00}.
For $\alpha_s\approx0.5$ the result in \cite{ar00} gives 
$\frac{\eta}{s}{\,=\,}0.135$. A lower limit for the shear viscosity 
in infinitely strongly coupled $N=4$ SUSY YM theory and variations 
thereof was derived in \cite{po01}, exploiting the AdS/CFT 
correspondence: $\frac{\eta}{s}{\,\geq\,}\frac{1}{4\pi}=0.08$.
In kinetic theory, in the Boltzmann gas approximation, the relaxation time
is estimated as $\tau_\pi=2 \eta \beta_2 =2\eta\, \frac{3}{4p}$ \cite{IS79}.

For the initial energy density distribution in the transverse plane,
we used a Woods-Saxon parameterisation,
\begin{equation}
e(r) =\frac{e_0}{1+e^\frac{r-R}{a}},
\end{equation}
with  $R{\,=\,}6.4$~fm, $a{\,=\,}0.54$~fm. This is not very realistic,
but facilitates comparison with the results of \cite{MR04}. 
($e_0{\,=\,}aT_i^4$ is the central energy density at initial time 
$\tau{\,=\,}\tau_i$.) I show results for initial conditions 
$T_i{\,=\,}0.3$~GeV and $\tau_i{\,=\,}0.5$~fm/$c$, with zero initial
radial flow ($v_r(r,\tau_i){\,=\,}0$). 

For the non-ideal fluid, initial viscous pressures $\pi_{\rm ini}^{rr}$ and
$\pi_{\rm ini}^{\phi\phi}$ are required. Even though $v_r$ and its derivatives 
are zero initially, due to the Bjorken longitudinal motion the stress 
tensor is not zero: $\tau_i^2\sigma^{\eta\eta}{\,=\,}-\frac{2}{3\tau_i}$, 
$r^2 \sigma^{\phi\phi}{\,=\,}\frac{1}{3\tau_i}$. 
We assume that at initial time $\tau_i$, the
viscous pressure components are fully relaxed to the Bjorken scaling 
expansion values,
\begin{equation}
\pi^{rr}_{\rm ini}=r^2 \pi_{\rm ini}^{\phi\phi} 
= - \frac{\tau_i^2}{2} \pi_{\rm ini}^{\eta\eta} = \frac{2\eta}{3\tau_i}.
\end{equation} 

\begin{figure}[h]
\begin{minipage}{12pc}
\includegraphics[bb=30 230 520 735,width=12pc,clip=]{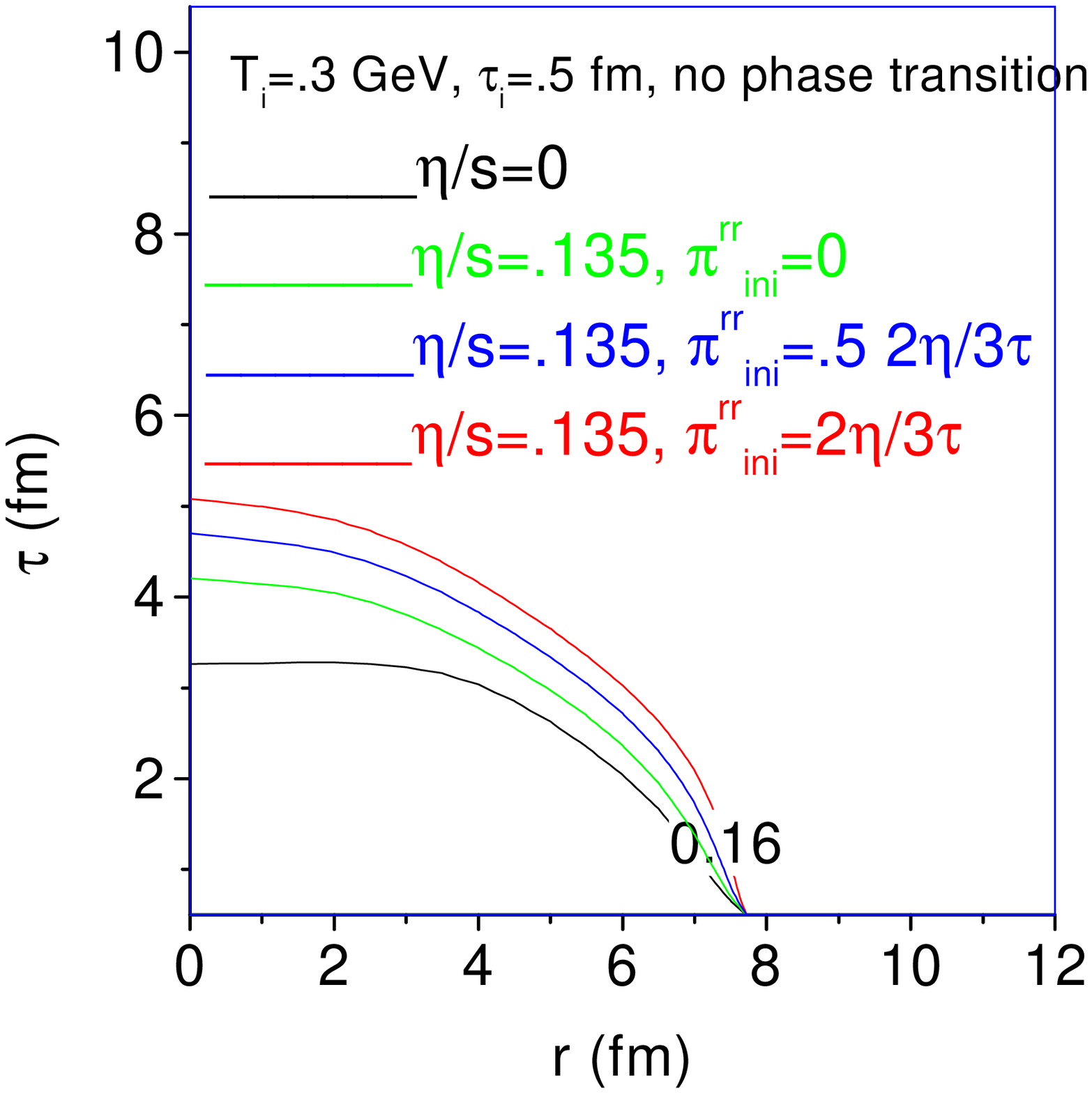}
\caption{\label{F3}
$T_f{=}160$\,MeV surface for different initial vis\-cous 
pressures.} 
\end{minipage}\hspace{1pc}%
\begin{minipage}{12pc}
\includegraphics[bb=30 230 520 735,width=12pc,clip=]{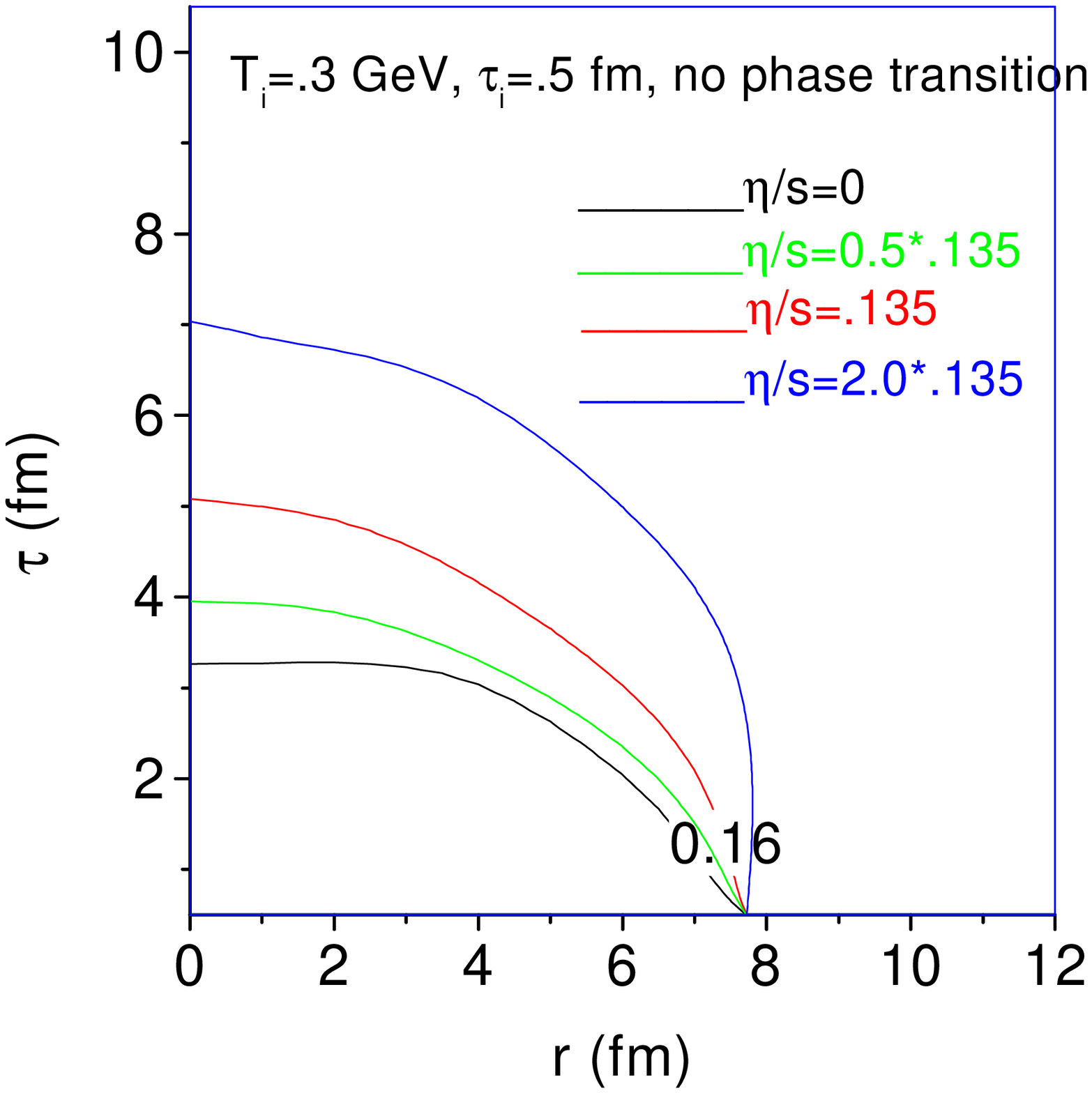}
\caption{\label{F4}
$T_f = 160$\,MeV surface for different shear viscosities
$\eta$.}
\end{minipage}\hspace{1pc}%
\begin{minipage}{12pc}
\includegraphics[bb=30 230 520 735,width=12pc,clip=]{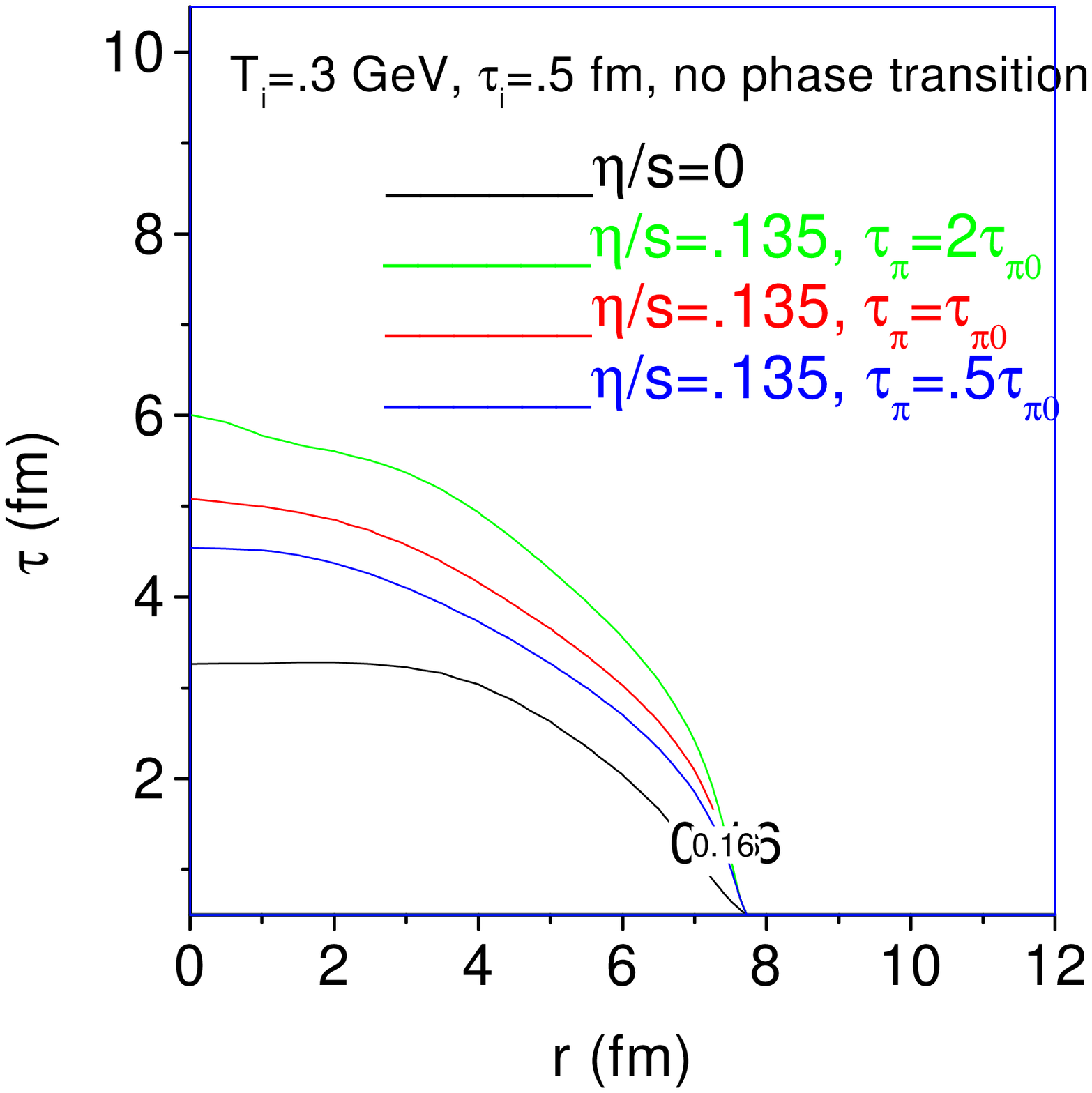}
\caption{\label{F5}
$T_f{=}160$\,MeV surface for different relaxation times
$\tau_\pi$.}
\end{minipage} 
\end{figure} 

Figures \ref{F3} to \ref{F5} show that the space-time evolution of 
non-ideal fluids depends sensitively on (i) the initial viscous 
pressure, (ii) the viscosity coefficient $\eta$, and (iii) the 
relaxation time $\tau_\pi$. In all three figures we show the shape
of the the $T_f{\,=\,}0.16$~GeV surface, with the solid black line 
giving the ideal fluid result for reference. Viscous effects are seen
to slow down the cooling of the matter, increasing the lifetime of the 
fireball and its average transverse size at freeze-out.  

In Fig.~\ref{F3} the $T_f{\,=\,}0.16$~GeV surface is shown for 
different initial viscous pressures,
$\pi_{\rm ini}^{rr}{\,=\,}0$, $\frac{\eta}{3\tau_i}$ and
$\frac{4\eta}{3\tau_i}$, respectively, using $\frac{\eta}{s}{\,=\,}0.135$. 
The higher the initial viscous pressure, the more extended is the 
freeze-out surface and the larger are the deviations from ideal fluid 
dynamics. The life time of the dissipative QGP is extended 
by 20\% if the initial viscous pressure is increased from zero to 
$\frac{4\eta}{3\tau_i}$. The freeze-out surface also depends 
sensitively on the value of the viscosity coefficient (Fig.~\ref{F4}). 
As the viscosity decreases, the departure of the freeze-out surface from 
ideal behavior also decreases. In Fig.~\ref{F5} we show the freeze-out 
surface for different relaxation times, $\tau_\pi=0.5\, \tau_{\rm kin}$, 
$\tau_{\rm kin}$ and $2\,\tau_{\rm kin}$ 
(where $\tau_{\rm kin}{\,=\,}\frac{3\eta}{2p}$), for fixed
viscosity $\eta/s{\,=\,}0.135$. As the relaxation time is increased 
by a factor 4, the freeze-out time in the fireball center increases 
by 25\%.

All the viscous effects shown in Figures \ref{F3} to \ref{F5} increase
if the initial time $\tau_i$ is decreased, keeping the total fireball 
entropy and all other parameters constant. This is due to the increasing 
initial longitudinal expansion rate $\theta{\,=\,}\frac{1}{\tau_i}$ which
results in a larger ratio $\frac{\tau_\pi}{\tau_i}$. This ratio is
the figure of merit which controls the importance of viscous corrections 
to ideal fluid dynamics. Our studies show that there is not only a minimum
thermalization time $\tau_{\rm therm}$ after which ideal fluid dynamic can 
be applied, but there is also a minimum time $\tau_i{\,<\,}\tau_{\rm therm}$
for the applicability of viscous fluid dynamics. The initial conditions
for viscous hydrodynamics at that time $\tau_i$ must be obtained by
matching the decomposition (\ref{eq17}) of the energy-momentum tensor
to the corresponding result for $T^{\mu\nu}$ from some preceding 
non-equilibrium kinetic evolution. 


\end{document}